\documentclass[aip,apl,reprint,twocolumn,superscriptaddress]{revtex4-1}                                       

\usepackage{graphicx}
\usepackage{dcolumn}
\usepackage{amsmath}
\usepackage{bm}
\usepackage{upgreek}
\usepackage{color}
\usepackage{natbib}

\makeatletter

\newcommand{\Rmnum}[1]{\expandafter\@slowromancap\romannumeral #1@}
\makeatother

\begin{document}
\title {Residual strain in free-standing CdTe nanowires overgrown with HgTe}

\author{Maximilian Kessel}
\email{maximilian.kessel@ntnu.no}
\affiliation{Physikalisches Institut (EP3), Universit\"at W\"urzburg, Am Hubland, 97074 W\"urzburg, Germany.}
\affiliation{Current address: Center for Quantum Spintronics, Department of Physics, Norwegian University of Science and Technology, NO-7491 Trondheim, Norway. }
\author{Lukas Lunczer}
\affiliation{Physikalisches Institut (EP3), Universit\"at W\"urzburg, Am Hubland, 97074 W\"urzburg, Germany.}
\author{Nadezda V. Tarakina}
\affiliation{Physikalisches Institut (EP3), Universit\"at W\"urzburg, Am Hubland, 97074 W\"urzburg, Germany.}
\affiliation{Current address: Max Planck Institute of Colloids and Interfaces, Am Mühlenberg 1, 14476 Potsdam, Germany.}
\author{Christoph Br\"une}
\affiliation{Physikalisches Institut (EP3), Universit\"at W\"urzburg, Am Hubland, 97074 W\"urzburg, Germany.}
\affiliation{Current address: Center for Quantum Spintronics, Department of Physics, Norwegian University of Science and Technology, NO-7491 Trondheim, Norway. }
\author{Hartmut Buhmann}
\affiliation{Physikalisches Institut (EP3), Universit\"at W\"urzburg, Am Hubland, 97074 W\"urzburg, Germany.}
\author{Laurens W. Molenkamp}
\affiliation{Physikalisches Institut (EP3), Universit\"at W\"urzburg, Am Hubland, 97074 W\"urzburg, Germany.}

\begin{abstract}
We investigate the crystal properties of CdTe nanowires overgrown with HgTe. Scanning electron microscopy (SEM) and scanning transmission electron microscopy (STEM) confirm, that the growth results in a high ensemble uniformity and that the individual heterostructures are single-crystalline, respectively. We use high-resolution X-ray diffraction (HRXRD) to investigate strain, caused by the small lattice mismatch between the two materials. We find that both CdTe and HgTe show changes in lattice constant compared to the respective bulk lattice constants. The measurements reveal a complex strain pattern with signatures of both uniaxial and shear strains present in the overgrown nanowires.
\end{abstract}

\maketitle

\section{Introduction}
HgTe is a topological insulator both in two and three dimensions\cite{Bernevig2006, koenig2007, fu2007, bruene2011} and the suitability for high quality MBE growth\cite{bruene2011} means that HgTe is a very interesting material for the growth of complex topological heterostructures. In topological insulators the charge transport is carried by surface states, which becomes additionally interesting for a nanowire configuration.
HgTe can be grown epitaxially on CdTe due to the close match in lattice constants (0.3\% mismatch). The tensile strain induced by this lattice mismatch is instrumental in bulk layers, where it lifts the degeneracy of the $\Gamma_8$ bands and opens a band gap\cite{bruene2011,kuo1985,zhao2013} such that the toplogical insulator state can be accessed.\cite{fu2007} 
We showed, recently, that it is possible to grow CdTe-HgTe nanowire structures,\cite{kessel2017} which are expected to show some residual strain.\cite{ferrand2014} Such quasi one-dimensional topological insulator structures are predicted to be candidates for Majorana states\cite{cook2011,alicea2012} and for unusual quantization features due to the properties of the topological surface states.\cite{egger2010}
In this work we investigate the crystal properties of such CdTe-HgTe nanowires using TEM and HRXRD. We focus on the signatures of strain in the heterostructures to establish if the ingredients for topological insulator behavior are present in the quasi one-dimensional HgTe crystallites.

\section{Methods}
CdTe NWs are grown by the vapor-liquid-solid (VLS) method on (110) GaAs substrates using a Au-Ga catalyst droplet and a short growth start with ZnTe to initiate vertical wire growth.\cite{kessel2017} They grow oriented uniformly along [111]B. In this nomenclature [111]B is pointing perpendicular out of a Tellurium terminated lattice plane with cubic close packing. Hence, (11$\overline{2}$)A are group-II terminated sides of the NWs. Subsequently, the (11$\overline{2}$)A sides of the CdTe NWs are overgrown with HgTe MBE. From electron diffraction experiments we know that the interface of both materials is formed by (11$\overline{1}$)A facets together with surface steps.\cite{kessel2017}\\\\
Transmission electron microscopy (TEM) studies were carried out by a FEI Titan 80-300 (S)TEM operated at 300\,kV. The cross-sectional TEM specimens were prepared in two steps. First, the sample was covered with PMMA and we cut a lamella. Then we thinned down the sample to a thickness of about 100\,nm by focused ion beam milling using an FEI Helios Nanolab Dual Beam system. We used a 30\,kV Ga\textsuperscript{+} ion beam with currents between 0.28\,nA and 28\,pA. A final low-kV cleaning was performed with a 2\,kV Ga\textsuperscript{+} ion beam and an 1$^{\circ}$ incident angle. High-angle annular dark-field scanning transmission electron microscopy (HAADF-STEM) images were recorded along [$\overline{1}$10] zone-axis with an electron energy of 300\,keV.\\\\
High-resolution X-ray diffraction (HRXRD) can be used to probe the crystalline structure of an ensemble of NWs in reciprocal space, similarly to a single crystal, provided the wires have identical structure and are all well orientated relative to the substrate. The diffraction pattern from the GaAs wafer is a reference to verify the sample alignment, which allows for an accuracy of less than 0.01\% relative error for the lattice constant derived from various reflections of the substrate. A detector array with spatial resolution of the diffracted X-ray wave is used to form an effective aperture, providing a high resolution along $2\theta$, the angle between primary beam and detector. At the same time, it is appropriately widened in $\omega$-direction, which is the angle between the beam and the substrate, in order to improve the signal-to-noise ratio. We investigate several reflections (33$\overline{5}$, 33$\overline{3}$, 33$\overline{1}$, 440, 331, 333 and 335) in the plane of NW's growth axis and substrate normal. 

\section{Results and discussion}
\subsection{Transmission Electron Microscopy results}
\begin{figure}[ht]
\begin{center}
		\includegraphics{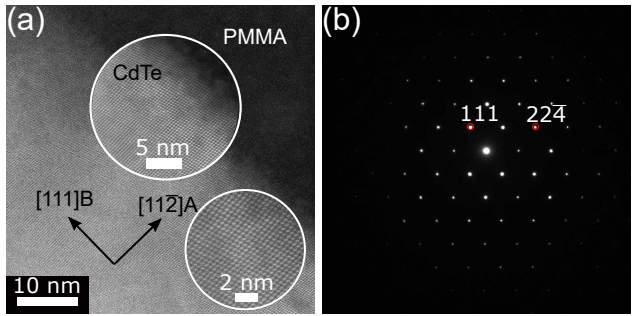}
\end{center}
		\caption{(a) HAADF-STEM images of a CdTe NW along [1$\overline{1}$0] zone axis and (b) the corresponding electron diffraction pattern.}
		\label{SEM_TEM_core}
\end{figure}
\begin{figure}[ht]
\begin{center}
		\includegraphics{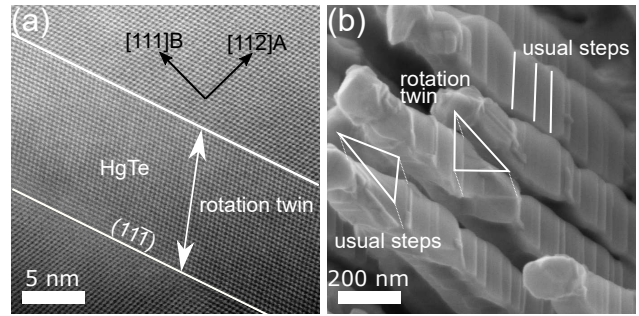}
\end{center}
		\caption{(a) HAADF-STEM image showing a rotation twin domain inside HgTe. (b) The rotation twin alters the shape.}
		\label{SEM_TEM_twin}
\end{figure}

We analyze the HAADF-STEM images for CdTe NWs growing along the [111]B direction of a (110) GaAs substrate. Figure~\ref{SEM_TEM_core}\,(a) shows HAADF-STEM images in the [1$\overline{1}$0] zone axis with higher magnifications as circular insets. The NW is single-crystalline and free of stacking faults over the entire length. Bright spots on HAADF-STEM images correspond to the position of the Cd and Te atoms. Since the atomic numbers of these elements are very close, 48 and 52, respectively, they can not be distinguished in the images. Figure~\ref{SEM_TEM_core}\,(b) shows the corresponding electron diffraction pattern which can be indexed in a cubic lattice, space group F$\overline{4}$3m, with an unit cell parameter $a\approx0.65$\,nm. This symmetry and lattice constant is consistent with zinc blende CdTe. Both methods, the real and the reciprocal space image in Fig.\,\ref{SEM_TEM_core}, show the high crystalline quality of the CdTe NWs.\\
\begin{figure}[ht]
\begin{center}
		\includegraphics{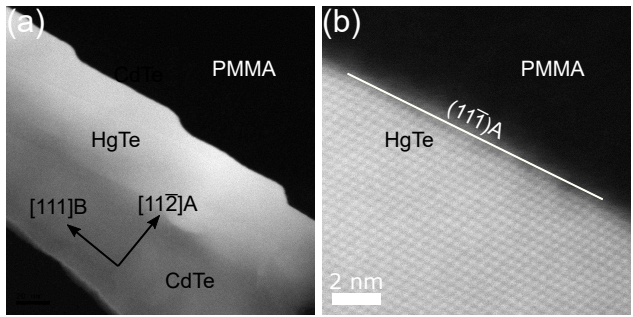}
\end{center}
		\caption{Image (a) shows the cross-section of HgTe grown on one side-facet of a CdTe NW. The HAADF-STEM image in (b) shows the crystal structure of the HgTe.}
		\label{SEM_TEM_shell}
\end{figure}
\\
The HAADF-STEM image in Fig.~\ref{SEM_TEM_shell}\,(a) shows an exemplary CdTe wire covered with epitaxial HgTe. The HgTe crystal grows predominantly on the (11$\overline{2}$)A side-facet of the CdTe NW, which is a result of sample-source geometry inside the growth chamber and different growth rates on facets with opposite polarity.\cite{kessel2017} Before overgrowth the (11$\overline{2}$) surface can be described by (11$\overline{1}$) facets and steps with a step length to height ratio of three to one.\cite{kessel2017} HgTe develops a pronounced (11$\overline{1}$) surface structure with steps visible in the HAADF-STEM and SEM images in Figs.\,\ref{SEM_TEM_shell} and \ref{SEM_TEM_twin}. These (11$\overline{1}$)A facets are a consequence of a step-flow growth mode with step bunching for HgTe on the (11$\overline{2}$)A CdTe side-facet. The high resolution HAADF-STEM image in Fig.~\ref{SEM_TEM_shell}\,(b) shows the HgTe zinc blende lattice along the [1$\overline{1}$0] zone axis. We find that the high CdTe crystal quality allows for the growth of epitaxial HgTe of comparable quality on each NW. In contrast to the segmented and polycrystalline HgTe-based NWs with inclusions of elemental Tellurium in earlier contributions,\cite{selvig2006,haakenaasen2008} our NWs show the desired zinc blende lattice of the II-VI material over the entire heterostructure. The results published earlier for a similar CdTe-based heterostructure show randomly oriented wires with respect to the substrate. \cite{wojnar2017} In contrast our samples have a high uniformity which allows to analyze the crystal structure with X-ray diffraction.\\
\\
The resolution of TEM is not enough to reveal the effects of the lattice mismatch, but it does allow for an analysis of grain boundaries in the NWs. Overgrowth occurs on the (11$\overline{1}$)A facets and twins with twin plane (11$\overline{1}$) are observed. Rotational twinning of 60$^\circ$ (or equally 180$^\circ$) around [11$\overline{1}$] is a possible defect in this growth mode.\cite{oron1988} This defect can be described as a local switching from cubic close packing (ccp) to hexagonal close packing (hcp) for one monolayer. Figure~\ref{SEM_TEM_twin}\,(a) shows a HAADF-STEM image of a rotation twin. The orientation of the crystal switches back to the original configuration after several atomic layers. The twin extends from the CdTe to the outer surface of the HgTe layer. These twins alter the orientation of stable facets and thereby the shape of the heterostructures, denoted in Fig.~\ref{SEM_TEM_twin}\,(b). The twinned region appears as a triangle, being rotated by 60$^\circ$ compared to the usual step configuration.
\subsection{X-ray diffraction results}
\begin{figure}[ht]
\begin{center}
	\includegraphics{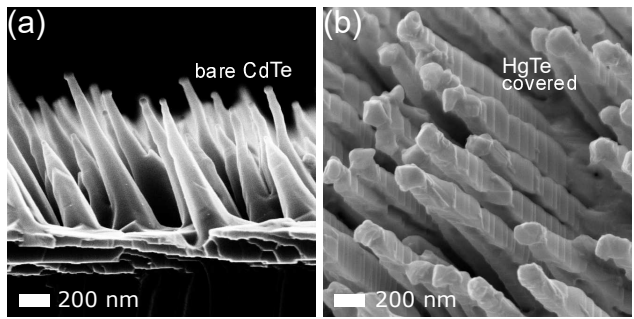}
\end{center}
		\caption[]{The SEM image in (a) shows bare CdTe NWs on GaAs (110) and (b) depicts a part of the same sample overgrown with HgTe.}
		\label{SEM_TEM_overgrowth}
\end{figure}
\begin{figure}[ht]
\begin{center}
		\includegraphics[width=\linewidth]{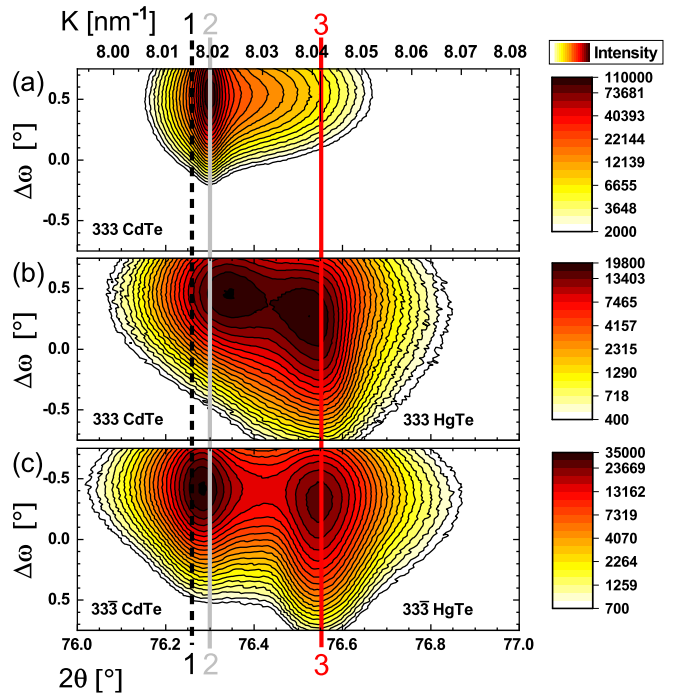}
\end{center}
		\caption{(a) 333 reflection of the bare CdTe NWs. (b) 333 and (c) 33$\overline{3}$ reflections of the NWs overgrown with HgTe. The black dashed line (1) gives the reference position for pure CdTe and the grey line (2) the position before overgrowth. The red line (3) gives the expected position for relaxed HgTe.}
		\label{RSM}
\end{figure}
\begin{figure}[h!]
\begin{center}
\includegraphics[width=\linewidth]{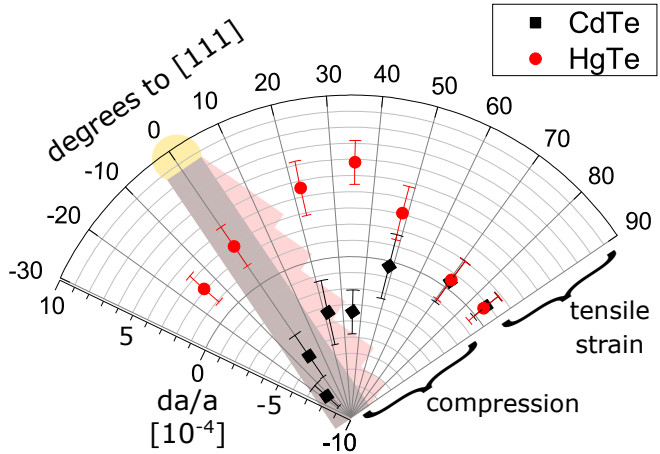}
\end{center}
		\caption{Direction dependent relative change of the plane spacing by residual strain in CdTe (black) NWs overgrown with HgTe (red).}
		\label{strain}
\end{figure}
\begin{figure}[ht]
\begin{center}
		\includegraphics[width=\linewidth]{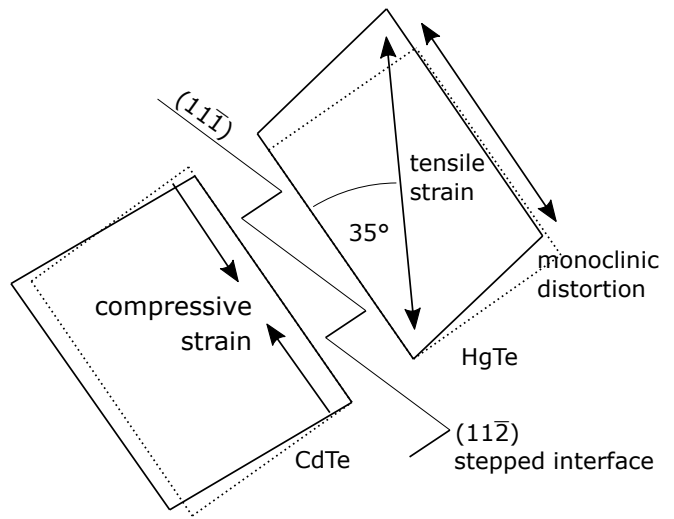}
\end{center}
		\caption[]{Sketch of the expected lattice distortion for mismatched epitaxy on a stepped interface. The relaxed unit cell in the [1$\overline{1}$0] zone axis is depicted by the dotted boxes for comparison.}
		\label{shear_strain}
\end{figure}

Residual strain analysis is performed for parallel oriented CdTe NWs overgrown with HgTe mainly on one side-facet. In this configuration, the superposition of X-rays diffracted by a collection of many wires can be treated like the diffraction pattern of a single-crystalline substrate with an epilayer.  A spatially integrating X-ray diffraction technique provides higher resolution in reciprocal space allowing us to resolve the influence of residual strain.  Figure~\ref{SEM_TEM_overgrowth} shows SEM micrographs of the samples investigated with HRXRD measurements, which are exemplarily shown in Fig.~\ref{RSM}. In order to analyze the evolution of strain due to the epitaxy of lattice mismatched materials we compare reflections of one sample before and after the overgrowth of HgTe, as shown in Figure~\ref{SEM_TEM_overgrowth}\,(a) and (b), respectively. To do so, some CdTe NWs are transferred out of UHV without being overgrown with HgTe. Then the sample is cleaved under N$_2$ atmosphere and one part is subsequently overgrown with HgTe.\\
We recorded reciprocal space maps (RSMs) for the bare CdTe NWs and the heterostructures, which allows us to identify the peaks corresponding to the CdTe and HgTe structures, respectively. Figure~\ref{RSM} shows representative RSMs of the 333 reflection from the bare CdTe NWs in (a) as well as 333 and 33$\overline{3}$ reflections recorded for the CdTe-HgTe structures in (b) and (c). The intrinsic lattice constant of HgTe is smaller than that of CdTe, thus we find the HgTe reflection at higher values of $2\theta$. The vertical lines in Figure~\ref{RSM} allow for a comparison to the expected values for a relaxed lattice. The dashed black line (1) gives the reference position for pure CdTe. We assumed a lattice constant $a_{\text{CdTe}}=0.6482(5)$\,nm, which was determined on many commercial CdTe substrates by HRXRD in our group. The grey line (2) denotes the position of the CdTe NW reflection before overgrowth.\\
\\
From the comparison of Fig.~\ref{RSM}\,(a) and (b) we find that the 333 reflection of the CdTe changes the position after HgTe overgrowth. This shift is the response of the lattice structure to the strain in the CdTe structure induced by epitaxy of HgTe with the aforementioned lattice mismatch. The lattice of CdTe is compressed by about 0.05\% parallel to the interface. The black data points in Fig.~\ref{strain} give the direction dependent relative deformation for the CdTe, due to HgTe overgrowth. The measured compression in the plane of the interface is expected for pseudomorphic epitaxy and it should be vice versa for the HgTe lattice.\\
In order to analyze the strain in the HgTe, we compare the measurements to a HgTe reference with zinc blende structure and lattice parameter $a_{\text{HgTe}}=0.6461(5)$\,nm. This value agrees with common literature data and is obtained by our group through fitting of HRXRD measurements of HgTe layers on CdTe wafers. The theoretical position of the reflections for a relaxed HgTe lattice is denoted with the red line (3) in Fig.~\ref{RSM}.\\
The red data points in Fig.~\ref{strain} depict the direction dependent relative deformation for HgTe compared to the reference lattice. The HgTe lattice is strained the most along [110], a direction tilted by 35$^\circ$ to the interface, which is stretched by about 0.06\%. The observed deformation of the unit cell is a consequence of shear strain induced by epitaxy of lattice mismatched materials and a stepped interface.\cite{Li1997}\\
\\
The experimental findings are sketched in Fig.~\ref{shear_strain}. A stepped interface strains the HgTe along two directions at once, resulting in a shear component. Additionally to the in-plane stretch parallel to the interface and the out-of-plane compressive response, a monoclinic distortion is visible. Shear strain leads to a further stretching of the lattice for directions close to [110], meaning around 35$^\circ$ to the interface.\\\\
The maximal direction dependent change of the lattice constant by strain is smaller than the nominal lattice mismatch of 0.3\%. Some part of the strain may relax elastically at the edges and corners of the small and free-standing NWs. Such surface and edge effects are known to play an important role for nanostructures.

\subsection{Influence of the ZnTe growth start}
In this section we discuss the shifted peak positions in the HRXRD measurements. The observed reflection in Fig.~\ref{RSM}\,(a) at slightly larger $2\theta$ for bare CdTe NWs (2) compared to the very pure reference material (1) is consistent for all investigated reflections. This can be explained by a very small Zn concentration throughout the CdTe nanowire, originating from the addition of Zn at the VLS growth start.\cite{kessel2017} By fitting various reflections and assuming Vegard's law we calculate an average Zn concentration $x=0.006$ in the Cd$_{1-x}$Zn$_{x}$Te. Although no Zn is applied for the actual CdTe wire growth, it remains solved in the catalytic droplet seeding the growth and gets slowly consumed during the ongoing crystal growth.\cite{kirmse2009}\\
In addition, we see that the $\omega$ position of the reflections in Fig~\ref{RSM} varies from the GaAs reference. We measure a tilt of 0.5$^{\circ}$ towards [001] for all reflections investigated. This can be explained by a bending of the NWs directly after the ZnTe growth start. Due to sample-source-geometry the materials are absorbed on one side of the catalytic droplet. The logical consequence is a lateral gradient in the Cd to Zn ratio in both the droplets and each individual NW shortly after switching the material sources. This results in a locally non-uniform lattice constant over the cross-section of the NWs, which builds up local shear strain bending the NWs at the ZnTe-CdTe interface.\\

\section{Conclusion}
Single-crystalline CdTe NWs with a high ensemble uniformity were covered by HgTe epitaxy on one side-facet. The heterostructures show the intended single-crystalline zinc blende structure over entire individual CdTe-HgTe nanowires. The lattice mismatch of the two materials causes residual lattice strain. We find about 0.05\% compression for CdTe and about 0.03\% stretching for HgTe parallel to their interface. The largest relative stretch of the HgTe lattice is 0.06\% measured along [110], 35$^{\circ}$ tilted to the interface. This monoclinic distortion is caused by shear strain related to a stepped interface. The result is in agreement with the stepped surface structure of CdTe NWs. The observation of strain in the CdTe-HgTe nanostructures is an important step towards the investigation of topological properties in further experiments. The strain lifts the degeneracy of the electronic bulk bands and opens a band gap such that the topological insulator state is established.

\section{acknowledgement}
This work was supported by the Deutsche Forschungsgemeinschaft in the priority progam ``Topological Insulators: Materials - Fundamental Properties - Devices'' (DFG-SPP 1666) and the Elitenetzwerk Bayern
(Doktorandenkolleg Topologische Isolatoren).


\bibliography{Ref}

\end{document}